\documentclass[a4paper]{jpconf}
\usepackage{graphicx}
\usepackage{wrapfig,lipsum,booktabs}
\usepackage{aas_macros}
\usepackage{amsmath}
\usepackage{amsfonts}
\usepackage{amssymb}

\usepackage{color}

\newcommand{\blue}{\textcolor{blue}}
\usepackage{hyperref}

\begin{document}
\pagenumbering{arabic}

\title{CORHEL-CME:  An Interactive Tool For Modeling Solar Eruptions}
\author{Jon A. Linker, Tibor Torok, Cooper Downs, Ronald M. Caplan, Viacheslav Titov, Andres Reyes, Roberto Lionello,  \& Pete Riley}
\address{Predictive Science Inc., 9990 Mesa Rim Road Suite 170, San Diego, CA  92121}
\ead{linkerj@predsci.com,tibor@predsci.com,cdowns@predsci.com,caplanr@predsci.com, titovv@predsci.com,areyes@predsci.com,lionel@predsci.com,pete@predsci.com}

\vspace{-1mm}

\begin{abstract}
Coronal Mass Ejections (CMEs) are immense eruptions of plasma and magnetic fields that are propelled outward from the Sun, sometimes with velocities greater than 2000 km/s.  They are responsible for some of the most severe space weather at Earth, including geomagnetic storms and solar energetic particle (SEP) events.  We have developed CORHEL-CME, an interactive tool that allows non-expert users to routinely model multiple CMEs in a realistic coronal and heliospheric environment.  The tool features a web-based user interface that allows the user to select a time period of interest, and employs RBSL flux ropes to create stable and unstable pre-eruptive configurations within a background global magnetic field.   The properties of these configurations can first be explored in a zero-beta magnetohydrodynamic (MHD) model, followed by complete CME simulations in thermodynamic MHD, with propagation out to 1 AU.  We describe design features of the interface and computations, including the innovations required to efficiently compute results on practical timescales with moderate computational resources.  CORHEL-CME is now implemented at NASA's Community Coordinated Modeling Center (CCMC) using NASA Amazon Web Services (AWS).  It will be available to the public by the time this paper is published. 

\end{abstract}

\section{Introduction}
\label{sec:intro}
Coronal Mass Ejections (CMEs) are magnificent displays of solar activity that propel plasma and magnetic fields outward into the solar wind.  In addition to the inherent scientific interest of such phenomena, CMEs play a crucial role in space weather at Earth (e.g., \cite{baker16}).  When the magnetic fields associated with CMEs contain significant southward pointed fields  (oppositely directed to  the magnetic field at the Earth's magnetopause, i.e., negative $B_z$ fields), they can reconnect with the Earth's magnetic field and peel away the protective magnetopause, an essential trigger to geomagnetic activity.  Therefore, the magnetic structure of CMEs is of vital interest.  The compression regions and shock waves created by these events also play an important role in the acceleration of solar energetic particles (SEPs) \cite{kahler92}.

CMEs have been observed for over four decades, and while a basic picture of their magnetic structure has emerged, fundamental questions remain.  
One of the key reasons why these questions remain unresolved is that it is difficult to measure the magnetic field in the corona.  CMEs produce ancillary signatures in a range of wavelengths that give important clues to the underlying structure of CMEs, such as EUV waves \cite{longetal2017}, EUV dimming signatures \cite{zhukov_auchere2004}, and the three-part structure of CMEs observed in white light \cite{vourlidasetal2013}.  In order to reveal the physical (e.g.,\,magnetic) structure responsible for these observed features, a model must include enough physics to reproduce observations for real events.  Thermodynamic magnetohydrodynamic (MHD) models of the solar corona,  which incorporate the relevant energy transport processes, have advanced to the point that they can predict EUV/X-ray emission \cite{lionello09a,downsetal2010,downsetal2013,vanderholstetal2014,mikic18,toroketal2018,sachdevaetal2023}, as well as scattering of white light observed in coronagraphs and at total solar eclipses, for direct comparison with observations. 
\begin{wrapfigure}{l}{0.45\textwidth}
  \centering
  \includegraphics[width=0.44\textwidth]{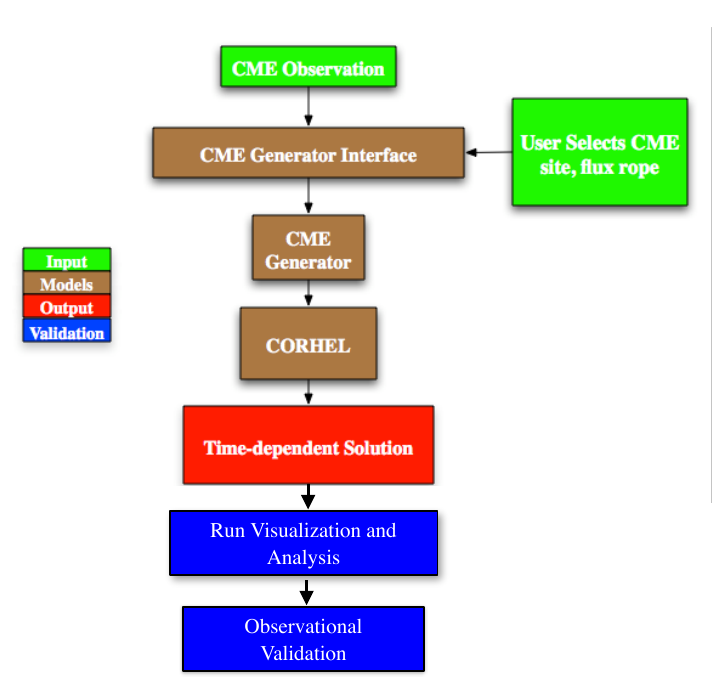}
\caption{Framework diagram for CORHEL-CME. 
	\label{fig_diagram} }
\end{wrapfigure}

Thermodynamic MHD models are now computed relatively routinely using the Magnetohydrodynamic Algorithm outside a Sphere (MAS) code within the CORona-HELiospheric (CORHEL) modeling suite.  However, it is still very challenging to model CME events, especially the most violent ones. The largest and fastest CMEs usually originate in localized active regions (ARs) on the Sun.  In addition to modeling the global corona, the detailed structure of AR must also be resolved.  To allow non-expert users to simulate CMEs in a realistic coronal and heliospheric environment, we have developed CORHEL-CME, which provides MAS/CORHEL zero-beta and thermodynamic MHD simulations using observed photospheric magnetic fields as boundary conditions.  Fig.\,\ref{fig_diagram} shows a diagram of the main steps in CORHEL-CME.  In this paper, we describe the software components of CORHEL-CME, the work flow for creating CME simulations, some of the computational shortcuts employed to make the CME simulations feasible on relatively modest computer hardware, and the standard diagnostics that are output as part of a run.  CORHEL-CME is now implemented at NASA's Community Coordinated Modeling Center (CCMC) using NASA Amazon Web Services (AWS).

\section{Components of CORHEL-CME}
Realistically simulating CMEs for a given event requires several key elements.  The background corona and solar wind must be simulated for the time period of interest; this in turn requires solar magnetic data.  This data is typically in the form of full-Sun maps of the line-of-sight magnetic field measured in the photosphere, from which the radial magnetic field is inferred.  Major CME events typically originate in solar active regions (ARs) where the magnetic fields are highly energized (energy above the potential state), and so a pre-eruptive configuration must be developed (also matching the photospheric field measurements).  To accomplish these tasks, CORHEL-CME employs the CORHEL modeling suite (described next), and models of pre-eruptive configurations based on analytical flux rope models (described in section \ref{sec_pec}).  Parameter specification  for CORHEL-CME is made via a Web interface (section \ref{sec_interface}.
\subsection{CORHEL}
\label{sec_corhel}
Advanced simulations of CMEs require a background model of the solar corona and solar wind, applicable to the time of the event that is being studied.  CORHEL is a suite of coupled models and tools for describing the 
corona and solar wind, developed by Predictive Science Inc. (PSI). It has been delivered to the CCMC and AFRL, and solutions are available at PSI's website (\blue{\href{http://www.predsci.com}{\path{www.predsci.com}}}).
The primary input data to CORHEL are synoptic maps of the radial photospheric magnetic field, $B_r$, from observatories such as the National Solar Observatories (NSO), SOLIS and GONG, and the Helioseismic and Magnetic Imager (HMI) aboard the Solar Dynamics Observatory (SDO).  An essential aspect of CORHEL's modeling approach is to divide the coronal and heliospheric domains, and to 
use different codes and approximations in each domain.  
The outer boundary for the coronal solutions is placed beyond the Alfv\'en and sound speed critical points, at 20--30 solar radii ($R_S$). The coronal solutions are used to 
drive the heliospheric solutions (extending to $\gtrsim$ 230$R_S$), which can be computed in either the inertial frame or a frame co-rotating with the Sun \cite{lionelloetal2013}. This approach has been applied to model the  propagation of observed CMEs from the low corona to 1 AU \cite{toroketal2018}. 
 The MHD runs use MAS---a high-performance code developed over several decades for studying the structure and dynamics of the global corona and inner heliosphere \cite{mikiclinker96,linker99a,mikic99a,riley01a,linkeretal2011,rileyetal2012,downsetal2013}. MAS can be run on massively parallel CPU systems using MPI, and has been ported to GPUs using OpenACC \cite{caplanetal2019} and Fortran standard parallelism \cite{caplanetal2023}.  In thermodynamic MHD mode, MAS incorporates a realistic energy equation; accounting for anisotropic thermal conduction, radiative losses, and coronal heating. This allows plasma densities and temperatures to be computed with sufficient accuracy to simulate extreme ultraviolet (EUV) and X-ray emission observed from space \cite{lionello09a}. 
The equations solved in MAS are shown in Appendix A of \cite{toroketal2018}.   The more recent versions of MAS employ a Wave-Turbulence-Driven (WTD) approach for coronal heating and solar wind acceleration;  \cite{downs16,mikic18} describe the additional equations solved.  To rapidly explore the stability and eruptive behavior of energized configurations, CORHEL-CME also employs the  simpler $\beta=0$ model (solution of the momentum equation and Faraday's law with plasma pressure = 0, e.g.,\cite{mikic94a}), prior to employing the more computationally intensive thermodynamic MHD model.

\subsection{CME Modeling}
MHD simulations are a powerful tool for studying CMEs.  Versions of the MAS code have been employed in state-of-the-art CME models for many years \cite{mikic94a,linker01a,linker03a,riley03a,riley07a,riley08b,lionelloetal2013,toroketal2018,downs21}. Such simulations are crucial for understanding CME initiation and evolution.  Traditionally they  have been time-consuming to develop and require significant expertise.  The goal of CORHEL-CME is to provide routine CME models that can be used by the broader scientific community.  The tool has been designed to be relatively straightforward to use, but it also incorporates enough details to produce sufficiently accurate results, including  proper modeling of the pre-eruptive state and early eruption phase. The latter is particularly important if the simulations are to be used in modeling of Solar Particle Events (SPEs), since compressional fronts, formed by the expansion of the CME, can accelerate particles at heights as low as 1.3\,$R_\odot$ in the corona \cite{kozarev16}.  A key component of CME simulations is specifying a pre-eruptive configuration, described in the next section.  CORHEL-CME is designed to allow the user to first investigate eruptions with the $\beta=0$ model, and, when the initial eruption properties are satisfactory, perform full Sun-to-1 AU simulations of the CME with the thermodynamic MHD model.

\subsubsection{Pre-Eruptive Configurations}
\label{sec_pec}
CORHEL-CME allows users to interactively perform routine simulations of CMEs based on realistic pre-eruptive conditions.  The TDM (Titov-D\'emoulin modified) model \cite{titovetal2014} is an extension of the TD model \cite{titov99a}, which has been widely used to study pre-eruptive magnetic configurations (e.g., \cite{kliem04})

and as the initial condition for CME simulations (e.g., \cite{torok05a,manchester08,lugaz09a}). The TD model is an approximately force-free equilibrium consisting of a toroidal MFR (partly submerged below the photosphere) and a stabilizing (`strapping') potential field produced by a fictitious pair of sub-photospheric monopoles. 
The TDm model employs the same MFR geometry, but was designed such that the MFR can be embedded into an arbitrary, locally bipolar potential field.
This facilitates the modeling of CMEs whose source regions have a relatively simple magnetic structure \cite{downs21}.  An earlier version of our tool, CORHEL-MAS-TDM, is available for runs on request at NASA CCMC 
(\blue{\href{https://ccmc.gsfc.nasa.gov/models/CORHEL~MAS-TDM~6.0}{\path{https://ccmc.gsfc.nasa.gov/models/CORHEL~MAS-TDM~6.0}}}) and provides $\beta=0$ modeling of CMEs in realistic solar magnetic fields.
To overcome the limitation of toroidal geometry, CORHEL-CME uses the RBSL (Regularized Biot-Savart Laws) model \cite{titov18,titov21}. This model contains an MFR of arbitrary shape, facilitating the modeling of complex CME source regions that have elongated and curved polarity inversion lines (PILs). Fig.\,\ref{fig_bastille} shows a schematic of the model and its application to the ``Bastille Day'' eruption that occurred on 7/14/2000.  Both CORHEL-MAS-TDM and CORHEL-CME allow the user to insert the MFRs into a given background field without altering the observed magnetogram.  

\subsection{Web Interface}
\label{sec_interface}
%
CORHEL-CME is delivered as two software packages:  One package is a GUI-based web interface, the other is the CORHEL modeling suite.  The web interface was developed with a client-server model using Django/Python and JavaScript.  It is installed on a public-facing machine (a NASA AWS instance in the CCMC implementation) and  allows users to interact with the interface via a web browser, eliminating the need for any local installation by users. The interface hides many details of CME modeling, allowing users to focus on a few free parameters that characterize pre-eruptive MFRs and their eruption.  Once all parameters are selected, CORHEL-CME creates two archive files for download. One file contains all information needed to repeat or modify a previous interface session; the other contains the files and scripts needed to run a simulation with CORHEL/MAS. The optimum specifics of the simulation (number of cores and grid points, computational mesh, run-time of the simulation) are determined automatically.

The CORHEL modeling suite is installed on the computer where the MHD computations are to be performed.  In the CCMC implementation, the archive files are automatically transferred to the computational node (at the present time, an 8-GPU NASA AWS Instance).  The CORHEL software  contains scripts and macros for automatic diagnostics, which are provided as a self-contained, interactive html page upon completion of the run (section \ref{sec_report}).

\section{CORHEL-CME Workflow}
\label{sec_workflow}
\begin{figure}[t]
\centering
\includegraphics[width=0.95\textwidth]{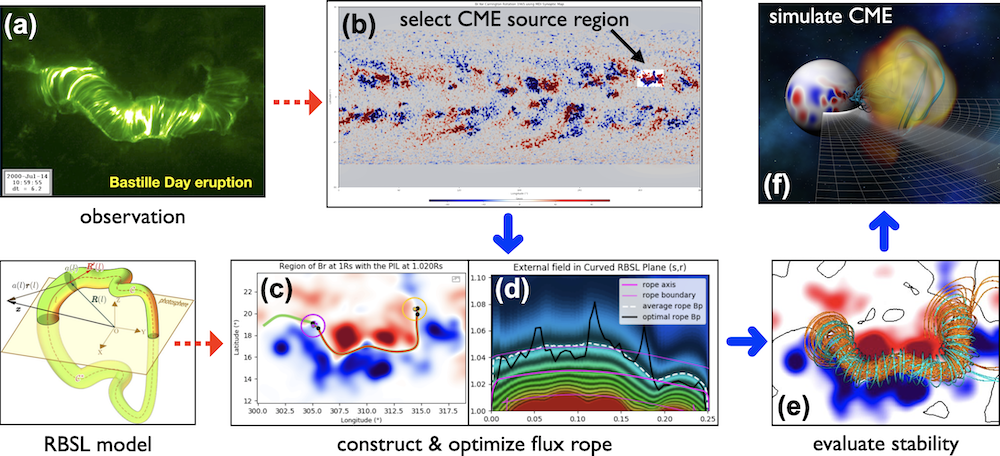}
\vspace{-1mm}
\caption{
Main steps of modeling a CME with CORHEL-CME. 
See text for details.
}
\vspace{-3mm}
\label{fig_bastille}
\end{figure}

\begin{figure}[t!]
\centering
\includegraphics[width=0.9\textwidth]{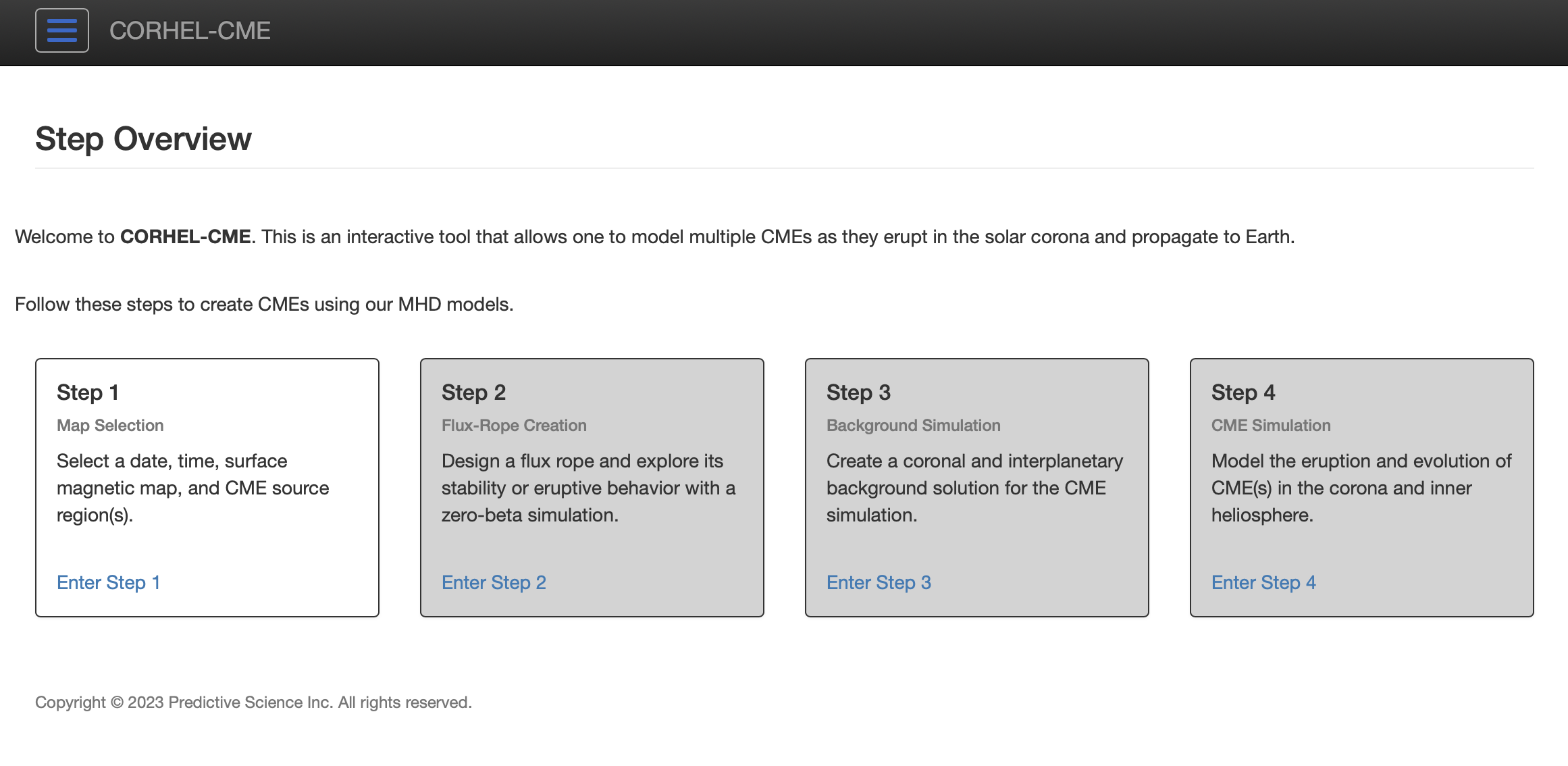}
\caption{
Navigation page of the CORHEL-CME web interface.  The user is guided through the process via various steps that include magnetogram and source-region selection (Step 1), RBSL flux-rope configuration (Step 2), Set-up of background models (Step 3) and final CME simulation (Step 4). 
}
\vspace{-3mm}
\label{fig_overview}
\end{figure}

Fig.\,\ref{fig_bastille} visually illustrates the workflow for creating a CME simulation with CORHEl-CME.  The user is guided through these steps via a navigation page, shown in Fig.\,\ref{fig_overview}.  First {\bf (STEP 1)}, the user selects the date and time of the CME to be modeled (Fig.\,\ref{fig_bastille}(a)).  The user is provided with choices for an appropriate surface magnetic map for the relevant Carrington rotation.  The user can also select an existing coronal and heliospheric background solution at this point, if such a solution is available in the database. In that case, the surface map from that solution is loaded and used in the subsequent steps.  After the data source is selected, it is then automatically downloaded from the corresponding data provider, and the user selects the location of the CME source region (typically an active region) on the map, as shown in Fig.\,\ref{fig_bastille}(b)).

Next {\bf (STEP 2)}, the user designs an RBSL flux rope that is inserted into the CME source region and provides the free magnetic energy required to power the CME (see Fig.\,\ref{fig_bastille}(c)).  This step includes (i) the selection of a segment of the polarity inversion line (PIL) along which the initial central section of the flux-rope axis shall be placed at an arbitrary height above the solar surface, (ii) selecting the locations of the rope footprints and its thickness, (iii) further fine-tuning of the axis geometry.  The user can modify the axis path, guided by a side-view of the strapping-field contours, as shown in Fig.\,\ref{fig_bastille}(d). Once all these parameters are chosen, a downloadable tar file is produced that contains all the necessary information and commands to run a zero-beta simulation on a supercomputer.  In the CCMC implementation, this file is automatically transferred to the compute server to perform the run.  The user can then perform computationally inexpensive $\beta=0$ MHD runs with varying field strengths to explore the stability of the MFR, to find the threshold for eruption, and to explore the initial CME trajectory.  These runs can typically be completed in 10-30 minutes of wall clock time on multi-core or GPU systems.  As interacting CMEs are an important phenomena, CORHEL-CME allows a user to implement multiple eruptions in multiple ARs, by proceeding through the steps repetitively.  Performing these types of calculations can greatly increase the computational requirements, so the CCMC implementation presently limits this capability to two events in a single run.

{\bf STEP 3} provides the capability to produce a background solution for the corona and inner heliosphere (technically, these are two separate calculations). These solutions are computed for individual Carrington rotations, and are required for the Sun-to-1 AU CME simulation (Step 4). These background calculations are intentionally performed with a relatively low spatial resolution, and are later re-meshed to high resolution for CME runs (see section \ref{sec_efficiency}).  In the present version, the user can choose between two empirical approaches for coronal heating (similar to \cite{lionello09a}), which are briefly explained in the corresponding interface tool tips.  We expect this to be updated with the WTD heating model (see section \ref{sec_corhel}) in the near future.  The heliospheric background calculation is fully automated and requires no user input.  
After the background run finishes, the background solution (both corona and heliospheric) is added to the database, which allows subsequent users to employ it in future calculations.  

Finally ({\bf STEP 4}), the user can perform a full Sun-to-1 AU CME run.  This requires a background solution for the Carrington rotation (corresponding to the time when the CME occurred), to be present in the CORHEL-CME background-solution database. The user will then select this solution during Step 1. If no such solution is present, the user has to perform Step 3 first before he/she can proceed further. Once a background solution is selected, the user continues to Step 2 to either construct a new flux rope or to ``confirm'' a previously constructed one (which can be loaded into the interface by restoring the corresponding session), after which he/she can directly go to Step 4 and start the CME simulation. No further input is required in Step 4.  Upon completion of each of these runs (zero-beta, background, CME), a tar file that contains an extensive run report (see Section\,\ref{sec_report}) is automatically generated and can be downloaded by the user. 

\section{Efficiency Considerations}
\begin{wrapfigure}{l}{0.51\textwidth}
  \centering
  \includegraphics[width=0.5\textwidth]{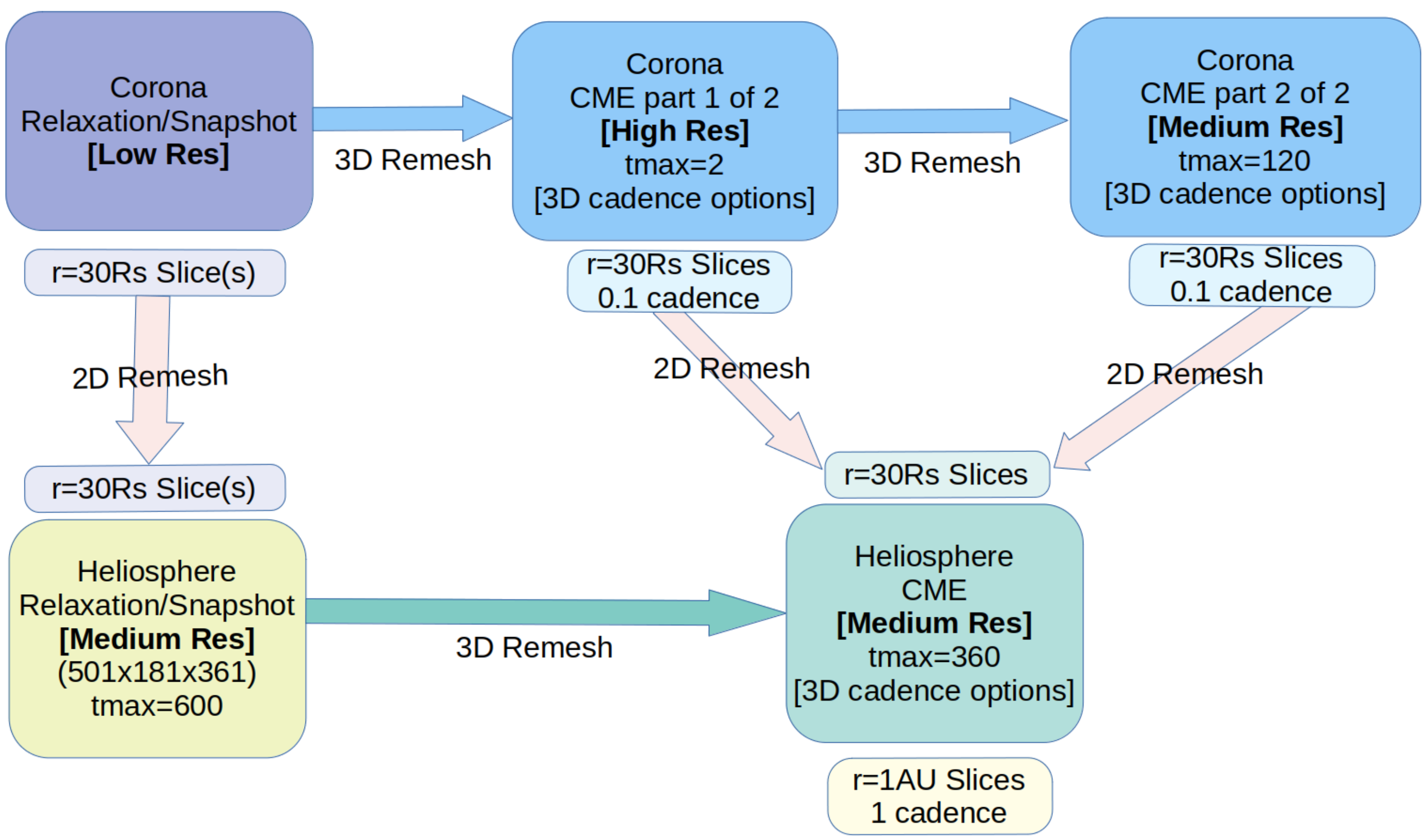}
\caption{Flow chart showing the run pipeline for modeling a single 
Sun-to-1 AU CME in the corona and inner heliosphere.  This 
corresponds to STEP 3 and STEP 4 in  Fig.\,\ref{fig_overview}. 
	\label{fig_pipeline} }
\end{wrapfigure}

\label{sec_efficiency}

An important goal for CORHEL-CME is to allow for real-world calculations with moderate computational resources (presently implemented on an 8-GPU card NASA AWS instance at CCMC).  This requires optimizing the required steps in the calculations. 
The sequence of MHD calculations performed, referred to as the run pipeline, is depicted in Fig.\,\ref{fig_pipeline}.

Global coronal simulations require days of physical time to approximate the corona, but only demand a few hours of computational time if ARs are resolved coarsely.  On the other hand, detailed models of the pre-event energized magnetic structure in the AR require high resolution to account for specific local features, where the magnetic field is very strong.  The local physical relaxation time in the AR is short (minutes to hours) because the Alfv\'en speed is very high in these magnetically dominated regions,  These considerations have necessitated  performing different steps of the CORHEl-CME pipeline with different resolutions and flexibly transitioning computed solutions between meshes of different resolution.  We interpolate the simulation variables with piece-wise monotonic splines (PCHIP) and analytic integral preserving functions for the transition from one mesh to the other (a special procedure is used to obtain the vector potential $\mathbf{A}$, see Appendix A).  We refer to this as re-meshing.  
\begin{figure}[h!]
\centering
\includegraphics[width=0.7\textwidth]{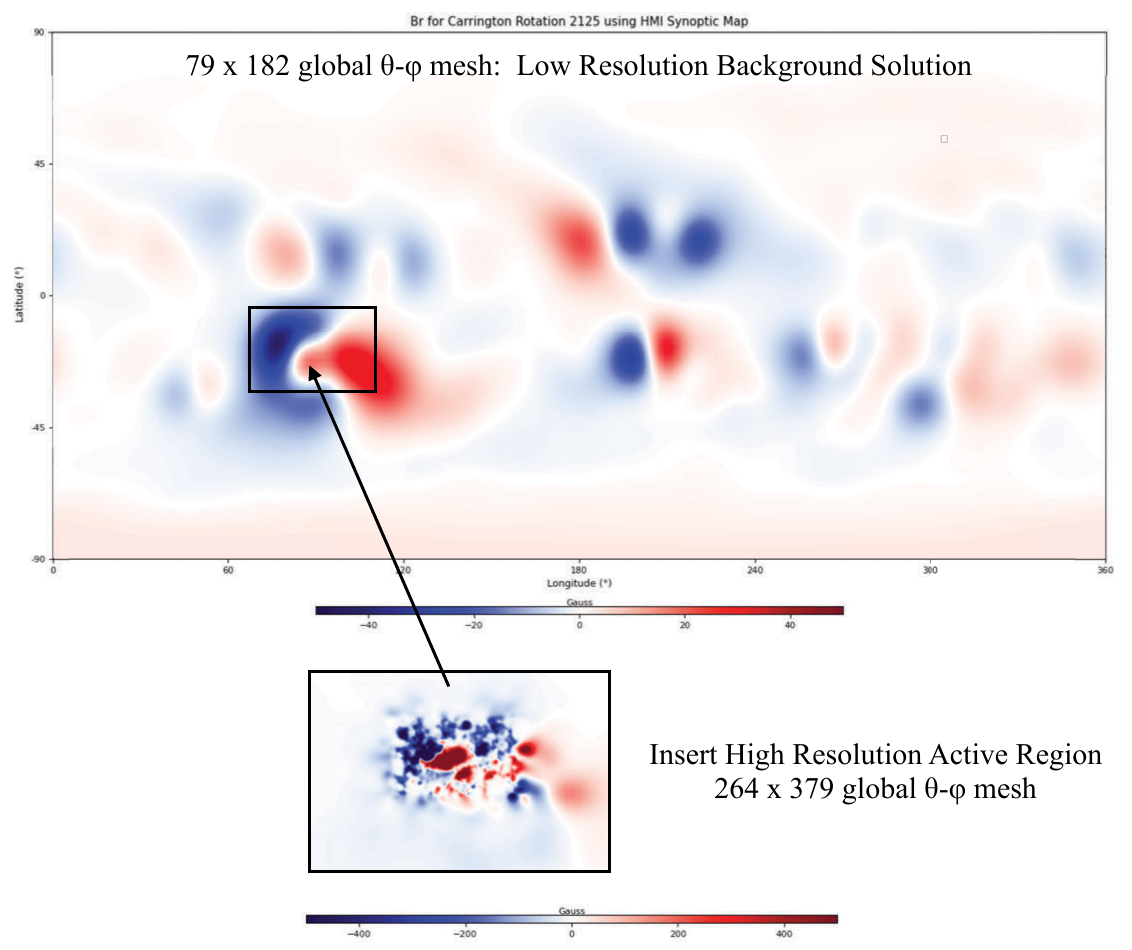}
\caption{
Latitude-longitude map of $B_r$ boundary for low resolution background simulation of the ambient corona around the time of the July 12, 2012 CME, together with the high resolution active region that would be inserted and re-meshed for a CORHEL-CME simulation.
}
\label{fig_remesh}
\end{figure}

Re-meshing is used at several points in the run pipeline, as shown in Fig.\,\ref{fig_pipeline} (a time t = 1 corresponds to 24 minutes).  For example, we compute the background simulation for the time period of interest at low resolution, to minimize run time and data storage.  If we performed the background simulation at the resolution required to resolve important features of the AR magnetic field, it greatly increases the computational requirements.  Instead, we interpolate our low-resolution simulation onto a higher-resolution grid, and insert the resolved active region into this background solution, prior to incorporating the RBSL flux rope (section \ref{sec_pec}).  (The $\beta=0$ flux-rope simulations are performed on the high resolution mesh.)  This procedure is depicted in Fig.\,\ref{fig_remesh}.  We have found that, in practice, the local perturbations introduced by this insertion resolve rapidly and do not greatly perturb the global coronal solution.  There is minimal effects on the subsequent propagation of the CME, as compared to a simulation that is performed entirely at the higher resolution.  We note that this re-meshing procedure differs from adaptive-mesh refinement approaches, because the {\it boundary condition} is modified; we are effectively solving a mathematically different problem from the one we started with.  We are relying on the physical effect that this change is local to a region and does not greatly influence the global solution.  We re-mesh again to an intermediate resolution after the initial eruption when the CME has strongly expanded, to compute the propagation of the CME until it fully exits the outer domain boundary of the coronal calculation at 30\,$R_\odot$.  The subsequent evolution is then followed in the heliospheric calculation, which is also performed more efficiently as a separate calculation.

\section{Diagnostics Report}
\label{sec_report}
Numerical simulations of solar eruptions are typically difficult to interpret, especially if they are performed with the thermodynamic MHD model.  CORHEL-CME incorporates the Psiweb Report Generator to create 
a self-contained interactive web report (as an HTML page), for a quick look and assessment of the MHD simulations.    The report generator sets up working directories for each report, gathers report parameters from configuration files, and launches PSI's post-processing tools written in Python, Bash, and Fortran. After all analyses are performed for a given report, a self-contained web document is generated using MkDocs (see \blue{\url{https://www.mkdocs.org/}}). 
Plotly's javascript library (see \blue{\url{https://plotly.com/javascript/}}) 
is used for three-dimensional (3D) field line exploration, while the 2D slices and emission imagery pages are driven with custom Javascript.  The report is automatically produced and tarred for download when a CORHEL-CME run has finished. 
\begin{figure}[h!]
\centering
\includegraphics[width=0.9\textwidth]{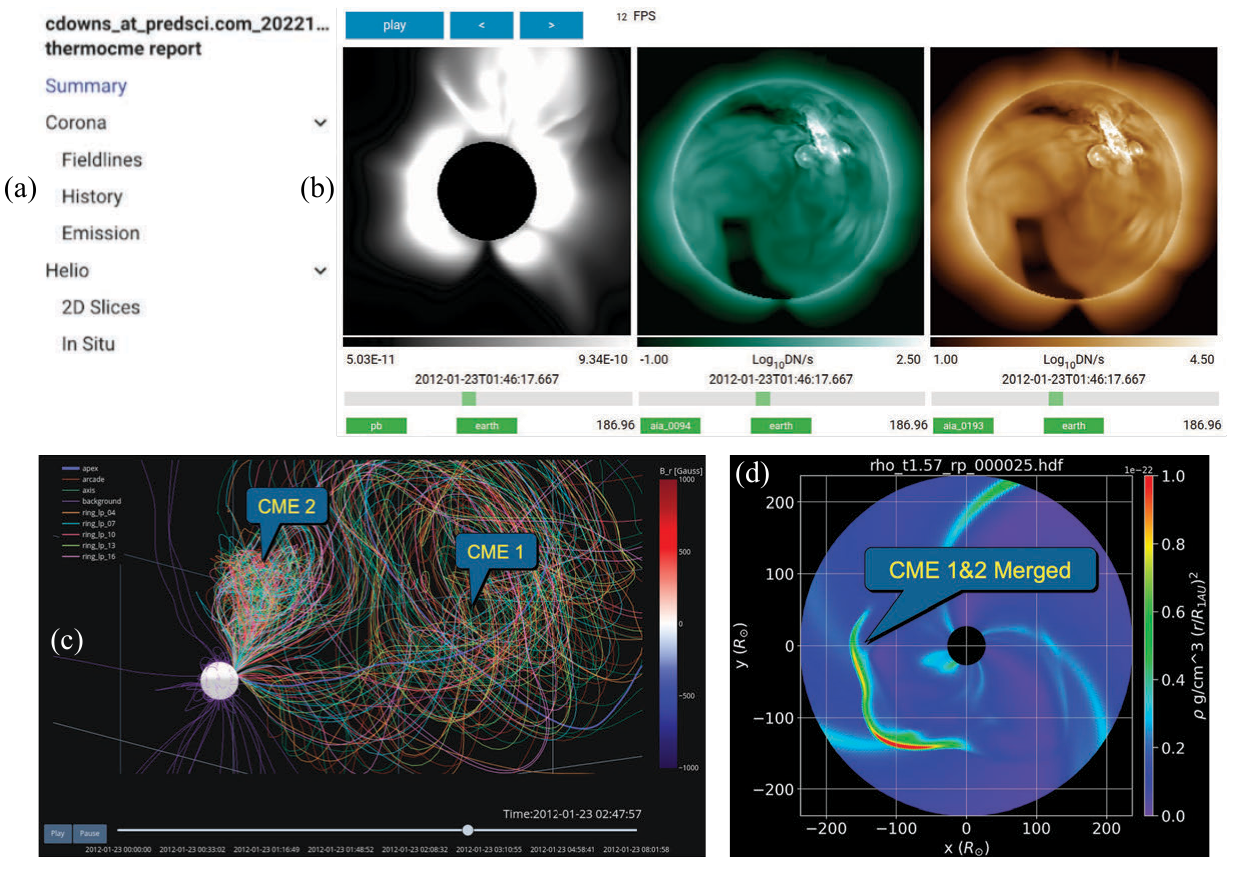}
\caption{
Field lines displayed by the CORHEL-CME interactive visualization tool, showing two flux ropes in the coronal simulation of the January 23, 2012 double-CME event.
The plasma density (scaled by the distance $r^2$ from the Sun) in the subsequent interplanetary simulation, shown in the equatorial plane from an interactive animation in the run report.}
\vspace{-3mm}
\label{fig_report}
\end{figure}

Fig.\,\ref{fig_report} shows excerpts from a report created for a simulation of the double CME event of January 23, 2012.  The report provides a drop-down menu so users can choose the set of variables animated on a given page (Fig.\,\ref{fig_report} (a).)  Movies of EUV and X-ray emission, as well as polarization brightness (pB)  as would be observed by different space-based instruments can be viewed from different vantage points in the ``emission'' page (Fig.\,\ref{fig_report}(b)).  The field-line tool (Fig.\,\ref{fig_report}(c)) allows users to plot different sets of field lines. The user can rotate the plot, zoom into an area, and follow the evolution of the field lines via a time slider.  The helio report (Fig.\,\ref{fig_report}(d)) provides an interactive animation of various physical quantities in the equatorial plane (30--230\,$R_\odot$).  Time series comparisons of the results with OMNI data can be obtained at the in situ page.

After completion of a run, the full MHD data from the simulation can also be obtained to perform more detailed analysis.

\section{Summary}
CORHEL-CME allows users to simulate CMEs for observed events, in coronal and heliospheric environments modeled with observed photospheric magnetic fields.  A GUI interface allows users to develop pre-eruptive configurations, and investigate stable flux ropes and eruptions in $\beta=0$ and full thermodynamic MHD.  After completion of a run, a downloadable diagnostics report is created with quick-look, interactive visualizations.  CORHEL-CME has been designed to operate with moderate computational resources, to allow more routine investigation of CME events.  More detailed simulations can subsequently be performed, if greater computational power is available.  This may be necessary for some types of scientific investigations.
CORHEL-CME is currently undergoing testing at NASA CCMC, implemented on NASA AWS.  It will be operational by the time of publication of this paper.   In the future, we will combine CORHEL-CME with the SPE Threat Assessment Tool (STAT, \cite{linkeretal2019,youngetal2021}) to allow users to simulate solar particle events.

\appendix
\section{}
The MAS MHD code solves for Faraday's law (advancement of the magnetic field), using the vector potential $\mathbf{A}$ ($ \mathbf{B} = \nabla \times  \mathbf{A}$).  In MAS, differential operators are formed on staggered meshes to ensure that standard vector identities (e.g $\nabla\cdot\nabla\times  = 0$) are preserved to round-off error.  Among other advantages, this enforces $\nabla\cdot \mathbf{B} = 0$ to machine precision in the calculations.  In re-meshing our solutions, the resampling of  $\mathbf{A}$ from coarse to fine meshes (or vice-versa),  can produce unacceptable errors, even when higher-order, smooth methods like PCHIP are used.  This occurs because the current density $ \mathbf{J} = \nabla \times (\nabla \times \mathbf{B})$ (normalized units), and the errors in $ \mathbf{J}$ directly impact the balance of forces.  To obtain the re-meshed $\mathbf{A^{RME}}$ with sufficient accuracy, we perform the following steps:

\begin{enumerate}
\item Remesh the original  boundary field $B_{r0}$ onto the new mesh, solve to obtain the  potential field $\mathbf{B^{pot0}}$.
\item Remesh to obtain the original $\mathbf{B}$ in the domain:  $\mathbf{B^{RM0}}$.
\item Obtain the nonpotential part of the original $\mathbf{B}$ in the domain: $\mathbf{B^{RME}} = \mathbf{B^{RM0}} - \mathbf{B^{pot0}}$.
\item Obtain the re-meshed $\mathbf{J^{RME}} = \nabla \times \mathbf{B^{RME}}$
\item Insert the  new boundary field ($B_{r1}$) for the new mesh (e.g. insert the high resolution $B_{r}$ at the boundary).  
\item Solve for the potential field to obtain $\mathbf{A^{pot1}}$ and $\mathbf{B^{pot1}}$ (MAS employs a potential solver that finds the vector potential for a $\nabla \times \mathbf{B} = 0$ potential field).
\item Utilize the following gauge choice for $\mathbf{A}$:

\begin{equation}
\label{Adef}
\mathbf{A} =
\left[ \begin{array}{c} A_r \\ A_{\theta} \\ A_{\phi}\end{array}\right]=\nabla \times
\left[ \begin{array}{c} \psi \\ 0 \\ 0\end{array}\right] + \left[ \begin{array}{c} A_r  \\ 0 \\ 0\end{array}\right] =
\left[ \begin{array}{c} A_r \\ \, 
\\ \dfrac{1}{r}\left(\dfrac{1}{\sin \theta}\,\dfrac{\partial \psi}{\partial \phi}\right) \\ \, 
\\ -\dfrac{1}{r}\,\dfrac{\partial \psi}{\partial \theta}
\end{array}\right].
\end{equation}
then 
\begin{equation}
\nabla^2_\perp A_r = - J^{{\rm RME}}_r; ~~~\nabla^2_\perp \psi = - B^{{\rm RME}}_r.
\end{equation}
\end{enumerate}
Equations (A.2) are series of 2D solves at each radial slice in the coronal domain, that can be performed rapidly with preconditioned conjugate gradient methods.  This determines the nonpotential portion of $\mathbf{A}$  ($\mathbf{A^{nonpot}}$).
The desired $\mathbf{A^{RME}} = \mathbf{A^{nonpot}} + \mathbf{A^{pot1}}$.
\section*{Acknowledgments}
This work was supported by AFOSR (contract FA9550-15-C-0001), the NASA SBIR program (contract 80NSSC19C0193), the NASA Living With a Star Strategic Capabilities program (grant 80NSSC22K0893), the NASA HTMS program (grant 80NSSC20K1274), the NSF PREEVENTS program (grant ICER1854790), the NASA HSOC program (grant 80NSSC20K1285), NASA grant 80NSSC23K0258, and the Parker Solar Probe WISPR contract NNG11EK11I to NRL (under subcontract N00173-19-C-2003 to PSI).  Computational resources were provided by the NASA High-End Computing (HEC) Program through the NASA Advanced Supercomputing (NAS) Division at Ames Research Center and by the Expanse supercomputer at the San Diego Supercomputing Center through the NSF ACCESS and XSEDE programs.

\section*{References}
\bibliographystyle{iopart-num}
\bibliography{corhel_cme}

\providecommand{\newblock}{}
\begin{thebibliography}{10}
\expandafter\ifx\csname url\endcsname\relax
  \def\url#1{{\tt #1}}\fi
\expandafter\ifx\csname urlprefix\endcsname\relax\def\urlprefix{URL }\fi
\providecommand{\eprint}[2][]{\url{#2}}

\bibitem{baker16}
{Baker} D~N and {Lanzerotti} L~J 2016 {\em American Journal of Physics\/} {\bf
  84} 166--180

\bibitem{kahler92}
Kahler S 1992 {\em Annual review of astronomy and astrophysics\/} {\bf 30}
  113--141

\bibitem{longetal2017}
{Long} D~M, {Bloomfield} D~S, {Chen} P~F, {Downs} C, {Gallagher} P~T, {Kwon}
  R~Y, {Vanninathan} K, {Veronig} A~M, {Vourlidas} A, {Vr{\v{s}}nak} B,
  {Warmuth} A and {{\v{Z}}ic} T 2017 {\em \solphys\/} {\bf 292} 7
  (\textit{Preprint} \eprint{1611.05505})

\bibitem{zhukov_auchere2004}
{Zhukov} A~N and {Auch{\`e}re} F 2004 {\em \aap\/} {\bf 427} 705--716

\bibitem{vourlidasetal2013}
{Vourlidas} A, {Lynch} B~J, {Howard} R~A and {Li} Y 2013 {\em \solphys\/} {\bf
  284} 179--201 (\textit{Preprint} \eprint{1207.1599})

\bibitem{lionello09a}
{Lionello} R, {Linker} J~A and {Miki{\'c}} Z 2009 {\em \apj\/} {\bf 690}
  902--912

\bibitem{downsetal2010}
{Downs} C, {Roussev} I~I, {van der Holst} B, {Lugaz} N, {Sokolov} I~V and
  {Gombosi} T~I 2010 {\em \apj\/} {\bf 712} 1219--1231 (\textit{Preprint}
  \eprint{0912.2647})

\bibitem{downsetal2013}
{Downs} C, {Linker} J~A, {Miki{\'c}} Z, {Riley} P, {Schrijver} C~J and
  {Saint-Hilaire} P 2013 {\em \sci\/} {\bf 340} 1196--1199

\bibitem{vanderholstetal2014}
{van der Holst} B, {Sokolov} I~V, {Meng} X, {Jin} M, {Manchester} W~B I,
  {T{\'o}th} G and {Gombosi} T~I 2014 {\em \apj\/} {\bf 782} 81
  (\textit{Preprint} \eprint{1311.4093})

\bibitem{mikic18}
{Miki{\'c}} Z, {Downs} C, {Linker} J~A, {Caplan} R~M, {Mackay} D~H, {Upton}
  L~A, {Riley} P, {Lionello} R, {T{\"o}r{\"o}k} T, {Titov} V~S, {Wijaya} J,
  {Druckm{\"u}ller} M, {Pasachoff} J~M and {Carlos} W 2018 {\em \natas\/} {\bf
  2} 913--921

\bibitem{toroketal2018}
{T{\"o}r{\"o}k} T, {Downs} C, {Linker} J~A, {Lionello} R, {Titov} V~S,
  {Miki{\'c}} Z, {Riley} P, {Caplan} R~M and {Wijaya} J 2018 {\em \apj\/} {\bf
  856} 75 (\textit{Preprint} \eprint{1801.05903})

\bibitem{sachdevaetal2023}
{Sachdeva} N, {Manchester} Ward~B I, {Sokolov} I, {Huang} Z, {Pevtsov} A,
  {Bertello} L, {Pevtsov} A~A, {Toth} G, {van der Holst} B and {Henney} C~J
  2023 {\em \apj\/} {\bf 952} 117 (\textit{Preprint} \eprint{2212.05138})

\bibitem{lionelloetal2013}
{Lionello} R, {Downs} C, {Linker} J~A, {T{\"o}r{\"o}k} T, {Riley} P and
  {Miki{\'c}} Z 2013 {\em \apj\/} {\bf 777} 76

\bibitem{mikiclinker96}
Miki\'c Z and Linker J~A 1996 {\em International Solar Wind 8\/} vol 382 ed
  Winterhalter D e~a (AIP Conf. Proceedings) p 104

\bibitem{linker99a}
Linker J~A, {Miki{\'c}} Z, Bisecker D~A, Forsyth R~J, Gibson S~E, Lazarus A~J,
  Lecinski A, Riley P, Szabo A and Thompson B~J 1999 {\em J. Geophys. Res.\/}
  {\bf 104} 9809

\bibitem{mikic99a}
Miki\'c Z, Linker J~A, Schnack D~D, Lionello R and Tarditi A 1999 {\em Phys.
  Plasmas\/} {\bf 6} 2217

\bibitem{riley01a}
Riley P, Linker J~A and Miki\'c Z 2001 {\em J. Geophys. Res.\/} {\bf 106} 15889

\bibitem{linkeretal2011}
{Linker} J~A, {Lionello} R, {Miki{\'c}} Z, {Titov} V~S and {Antiochos} S~K 2011
  {\em \apj\/} {\bf 731} 110

\bibitem{rileyetal2012}
{Riley} P, {Linker} J~A, {Lionello} R and {Mikic} Z 2012 {\em \jastp\/} {\bf
  83} 1--10

\bibitem{caplanetal2019}
{Caplan} R~M, {Linker} J~A, {Miki{\'c}} Z, {Downs} C, {T{\"o}r{\"o}k} T and
  {Titov} V~S 2019 {\em ASTRONUM\/} ({\em Journal of Physics Conference
  Series\/} vol 1225) p 012012 (\textit{Preprint} \eprint{1811.02605})

\bibitem{caplanetal2023}
Caplan R~M, Stulajter M~M and Linker J~A 2023 {\em 2023 IEEE International
  Parallel and Distributed Processing Symposium Workshops (IPDPSW)\/} pp
  582--590

\bibitem{downs16}
{Downs} C, {Lionello} R, {Miki{\'c}} Z, {Linker} J~A and {Velli} M 2016 {\em
  \apj\/} {\bf 832} 180 (\textit{Preprint} \eprint{1610.02113})

\bibitem{mikic94a}
Miki\'c Z and Linker J~A 1994 {\em \apj\/} {\bf 430} 898

\bibitem{linker01a}
Linker J~A, Lionello R, {Miki{\'c}} Z and Amari T 2001 {\em J. Geophys. Res.\/}
  {\bf 106} 25165

\bibitem{linker03a}
Linker J~A, {Miki{\'c}} Z, Lionello R, Riley P, Amari T and Odstrcil D 2003
  {\em Phys. Plasmas\/} {\bf 10} 1971

\bibitem{riley03a}
{Riley} P, {Linker} J~A, {Miki{\'c}} Z, {Odstrcil} D, {Zurbuchen} T~H, {Lario}
  D and {Lepping} R~P 2003 {\em J. Geophys. Res.\/} {\bf 108} 2

\bibitem{riley07a}
{Riley} P, {Lionello} R, {Miki{\'c}} Z, {Linker} J, {Clark} E, {Lin} J and {Ko}
  Y~K 2007 {\em \apj\/} {\bf 655} 591--597

\bibitem{riley08b}
{Riley} P, {Lionello} R, {Miki{\'c}} Z and {Linker} J 2008 {\em \apj\/} {\bf
  672} 1221--1227

\bibitem{downs21}
{Downs} C, {Warmuth} A, {Long} D~M, {Bloomfield} D~S, {Kwon} R~Y, {Veronig}
  A~M, {Vourlidas} A and {Vr{\v{s}}nak} B 2021 {\em \apj\/} {\bf 911} 118

\bibitem{kozarev16}
{Kozarev} K~A and {Schwadron} N~A 2016 {\em \apj\/} {\bf 831} 120
  (\textit{Preprint} \eprint{1608.00240})

\bibitem{titovetal2014}
{Titov} V~S, {T{\"o}r{\"o}k} T, {Mikic} Z and {Linker} J~A 2014 {\em \apj\/}
  {\bf 790} 163

\bibitem{titov99a}
{Titov} V~S and {D{\'e}moulin} P 1999 {\em \aap\/} {\bf 351} 707--720

\bibitem{kliem04}
{Kliem} B, {Titov} V~S and {T{\"o}r{\"o}k} T 2004 {\em \aap\/} {\bf 413}
  L23--L26 (\textit{Preprint} \eprint{arXiv:astro-ph/0311199})

\bibitem{torok05a}
{T{\"o}r{\"o}k} T and {Kliem} B 2005 {\em \apjl\/} {\bf 630} L97--L100
  (\textit{Preprint} \eprint{astro-ph/0507662})

\bibitem{manchester08}
{Manchester} IV W~B, {Vourlidas} A, {T{\'o}th} G, {Lugaz} N, {Roussev} I~I,
  {Sokolov} I~V, {Gombosi} T~I, {De Zeeuw} D~L and {Opher} M 2008 {\em \apj\/}
  {\bf 684} 1448--1460 (\textit{Preprint} \eprint{0805.3707})

\bibitem{lugaz09a}
{Lugaz} N, {Roussev} I~I and {Sokolov} I~V 2009 {\em IAU Symposium\/} ({\em IAU
  Symposium\/} vol 257) ed {N~Gopalswamy \& D~F~Webb} pp 391--398

\bibitem{titov18}
{Titov} V~S, {Downs} C, {Miki{\'c}} Z, {T{\"o}r{\"o}k} T, {Linker} J~A and
  {Caplan} R~M 2018 {\em \apjl\/} {\bf 852} L21 (\textit{Preprint}
  \eprint{1712.06708})

\bibitem{titov21}
{Titov} V~S, {Downs} C, {T{\"o}r{\"o}k} T, {Linker} J~A, {Caplan} R~M and
  {Lionello} R 2021 {\em \apjs\/} {\bf 255} 9 (\textit{Preprint}
  \eprint{2106.02789})

\bibitem{linkeretal2019}
{Linker} J~A, {Caplan} R~M, {Schwadron} N, {Gorby} M, {Downs} C, {Torok} T,
  {Lionello} R and {Wijaya} J 2019 {\em ASTRONUM\/} ({\em Journal of Physics
  Conference Series\/} vol 1225) p 012007 (\textit{Preprint}
  \eprint{1905.05299})

\bibitem{youngetal2021}
{Young} M~A, {Schwadron} N~A, {Gorby} M, {Linker} J, {Caplan} R~M, {Downs} C,
  {T{\"o}r{\"o}k} T, {Riley} P, {Lionello} R, {Titov} V, {Mewaldt} R~A and
  {Cohen} C~M~S 2021 {\em \apj\/} {\bf 909} 160 (\textit{Preprint}
  \eprint{2012.09078})

\end{thebibliography}

\end{document}